\documentclass[reprint, superscriptaddress, nofootinbib,aps,prl]{revtex4-1}

\pdfoutput=1
\usepackage{mathtools, amsfonts, amsthm, latexsym, amssymb,dsfont}
\usepackage[T1]{fontenc}
\usepackage{microtype}
\usepackage{times}

\usepackage{hyperref}
\hypersetup{colorlinks=true, citecolor=blue, urlcolor=blue}

\newcommand{\mO}{\mathcal{O}}

\newcommand{\pd}{\partial}

\newcommand{\nn}{\nonumber}

\usepackage{color}

\begin{document}
\title{Renormalization Group Flows on Line Defects}
\author{Gabriel Cuomo}
\email{gcuomo@scgp.stonybrook.edu}
\affiliation{Simons Center for Geometry and Physics, SUNY, Stony Brook, NY 11794, USA}
\affiliation{C. N. Yang Institute for Theoretical Physics, Stony Brook University, Stony Brook, NY 11794, USA}
\author{Zohar Komargodski}
\email{zkomargodski@scgp.stonybrook.edu}
\affiliation{Simons Center for Geometry and Physics, SUNY, Stony Brook, NY 11794, USA}
\affiliation{C. N. Yang Institute for Theoretical Physics, Stony Brook University, Stony Brook, NY 11794, USA}
\author{Avia Raviv-Moshe}
\email{araviv-moshe@scgp.stonybrook.edu}
\affiliation{Simons Center for Geometry and Physics, SUNY, Stony Brook, NY 11794, USA}
\date{\today}

\begin{abstract}
We consider line defects in $d$-dimensional Conformal Field Theories (CFTs). The ambient CFT places nontrivial constraints on Renormalization Group (RG) flows on such line defects. We show that the flow on line defects is consequently irreversible and furthermore a canonical decreasing entropy function exists. This construction generalizes the $g$ theorem to line defects in arbitrary dimensions. We demonstrate our results in a flow between Wilson loops in 4 dimensions.
\end{abstract}
\maketitle

\paragraph*{Introduction.} In lattice systems, in order to understand the physics on different length scales, we perform block-spin transformations, eliminating degrees of freedom that live at short distances. This process obviously reduces the overall number of degrees of freedom. But one can ask whether this reduces the number of degrees of freedom per lattice site, which is much less clear. 
In Quantum Field Theory, the number of degrees of freedom per lattice site is roughly speaking the number of fields and this raises the question of whether the number of fields decreases as we probe physics of longer and longer distances. 

To address these questions precisely one has to give a non-perturbative definition of what the ``number of fields'' means and provide a prescription to evaluate it even when there is no weakly-coupled description in terms of fields. Starting from the work of Zamolodchikov on the $c$-function in 2d~\cite{1986JETPL..43..730Z}, several such proposals and results were discussed in diverse dimensions \cite{Cardy:1988cwa, 
Cappelli:1990yc,Osborn:1991gm,Myers:2010tj, Jafferis:2011zi, Komargodski:2011vj, Casini:2012ei,Elvang:2012st,Elvang:2012yc,Yonekura:2012kb,Antipin:2013pya,Grinstein:2013cka,Jack:2013sha,Baume:2014rla,Grinstein:2014xba,Giombi:2014xxa,Jack:2015tka,Cordova:2015fha,Casini:2015woa, Pufu:2016zxm, Casini:2017vbe,Fluder:2020pym,Delacretaz:2021ufg}. 

The focus of this paper is the physics of 1 dimensional defects in a CFT. Such defects can undergo nontrivial renormalization group flows while affecting the bulk very little far away from the defect.  
A few known examples of this kind include Wilson or 't Hooft lines in 4d gauge theories \cite{Kapustin:2005py} and holography \cite{Gomis:2006sb,Polchinski:2011im},  symmetry defects and impurities in 3d quantum critical systems \cite{Billo:2013jda,Gaiotto:2013nva, Giombi:2021uae,sachdev1999quantum,vojta2000quantum,liu2021magnetic} etc.  In 2d, line defects correspond to boundaries or interfaces and appear naturally as the low-energy limit of lattice systems with impurities (see, for instance, \cite{tsvelick1985exact,Ishibashi:1988kg,Cardy:1989ir,Affleck:1992ng,Affleck:1995ge}). 

There is already some extensive work on renormalization group flows on various defects 
\cite{Affleck:1991tk, Dorey:1999cj,Yamaguchi:2002pa,Friedan:2003yc,Azeyanagi:2007qj,Takayanagi:2011zk,Estes:2014hka,Gaiotto:2014gha,Jensen:2015swa,Casini:2016fgb,Andrei:2018die,Kobayashi:2018lil,Casini:2018nym,Giombi:2020rmc,Wang:2020xkc,Nishioka:2021uef,Wang:2021mdq,Sato:2021eqo}. For our purposes, it is important to highlight the conjecture of Affleck and Ludwig~\cite{Affleck:1991tk} for the decreasing entropy function on line defects in 2 dimensions and its subsequent proofs~\cite{Friedan:2003yc} and \cite{Casini:2016fgb}. Here we will discuss the properties of line defects in arbitrary dimensions. We will define an entropy function and show that it monotonically decreases. 
In the Supplemental Material we show how our result applies to a nontrivial flow between two different conformal Wilson lines in super Yang-Mills (SYM) theory in 4 dimensions. 

The main idea we employ is that surrounding the line defect with conformal charges leads to nontrivial identifications in theory space when the defect is non-conformal. This can be expressed in terms of constraints on the dilaton living on the line defect. We show that these constraints translate to a monotonic entropy function.

\paragraph*{DCFTs} We consider local, reflection-positive Euclidean conformal field theories (CFTs) in $d\geq 2$ dimensions. We will be interested in CFTs in the presence of a line defect  which preserves  unitarity and locality. 
We will be interested in infinite straight lines or circular defects.  At the fixed point of the (defect) renormalization group flow, the straight line defect preserves the subgroup $SL(2,\mathbb{R})\times SO(d-1)$ of the full conformal group.
In this case the system is called a defect CFT (DCFT). In $d=2$, conformal line defects additionally preserve one copy of the Virasoro algebra. 

DCFTs share many of the standard properties of CFTs.  
However, in general, the line defect does not support a stress tensor \cite{Nakayama:2012ed,Billo:2016cpy,Herzog:2017xha}.
This statement really means that there is no possibility to localize energy on the line defect and energy always ends up being smeared into the bulk.
The bulk stress tensor $T^{\mu\nu}_b$ obeys the following Ward identity \cite{Osborn:1993cr, Jensen:2015swa, Billo:2016cpy,Cuomo:2021cnb}:\footnote{It is convenient to consider normalized correlation functions, so $\langle T^{\mu\nu}_b\rangle$ really stands for $\langle T^{\mu\nu}_bD\rangle/\langle D\rangle$ where $D$ is the defect operator.}
\begin{equation}\label{eq_Conservation_DCFT}
\nabla_{\mu}  T^{\mu\nu}_b =-\delta_D^{d-1}n^\nu_i D^i \,,
\end{equation}
where $\delta_D^{d-1}$ is a delta function localized at the defect, $\{n^\nu_i\}$ is a basis of $d-1$ unit vectors normal to the defect and $D^i$ is the displacement operator \cite{Jensen:2015swa,Billo:2016cpy},\footnote{The defect $D$, and the displacement operator, $D^i$,  are distinguished by the superscript $i$.} which parametrizes the breaking of translations in the directions normal to the defect.  Finally we mention that all bulk correlation functions may be systematically decomposed into defect correlators via the bulk-to-defect OPE \cite{Cardy:1989ir,Gliozzi:2015qsa}. This allows to study the DCFT data via a systematic bootstrap approach \cite{Liendo:2012hy,Billo:2016cpy, Lauria:2020emq}, similar to the one usually adopted in standard CFTs \cite{Belavin:1984vu,Rattazzi:2008pe,Poland:2018epd}.   

It will be convenient for our purposes to consider the expectation of the $SL(2,\mathbb{R})$ charges wrapping the defect. These are obtained by integrating the stress tensor contracted with the appropriate Killing vector at a fixed distance $\varepsilon$ from the defect:
\begin{equation}\label{eq_charge}
Q_\xi(D)=\int_{\varepsilon} d^{d-1}\Sigma^\mu\langle T^b_{\mu\nu}\rangle\xi^\nu\,.
\end{equation}
By conformal invariance we expect eq. \eqref{eq_charge} to yield a vanishing result for both a straight line defect and a circular one. However, due to a subtlety with the action of conformal transformations on the point at infinity,  for the straight line geometry the conformal charges vanish only when the distance between the integration surface and the defect diverges sufficiently fast as $x^d\rightarrow\pm \infty$ \cite{Kapustin:2005py}. This issue is presumably related to the disagreement between the expectation value of circular and linear Maldacena-Wilson loops in $\mathcal{N}=4$ SYM \cite{Erickson:2000af, Drukker:2000rr,Pestun:2007rz}.  We provide a detailed discussion regarding this subtlety in the Supplemental Material. No issues of this sort arise for circular defects, hence we will focus on this geometry in what follows.

\begin{figure}[t!]
\centering
\includegraphics[width=55mm]{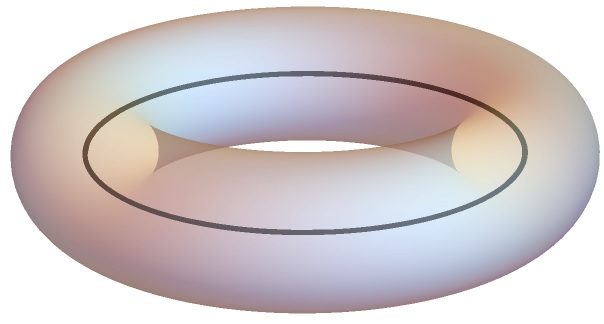}
\caption{An illustration of a toroidal surface wrapping a circular defect. \label{TorusFig}}
\end{figure}

Let us consider for concreteness a circular defect of radius $R$ centered around the origin on the $(x^1,x^2)$ plane $x^3=\ldots=x^d=0$. The $SL(2,\mathbb{R})$ Killing vectors preserved by the circle are: 
\begin{align}\nn
\xi_{(a)}^\mu&=\frac12\left[\delta_a^\mu
\left(R+x^2/R\right)-2x^\mu x_a/R\right]
\,,\\
\xi_{(\phi)}^\mu&=\delta^\mu_a\epsilon^{ab}x_b\,, \label{eq_Killing}
\end{align}
where $a=1,2$ and indices are raised/lowered with the Euclidean metric. Here $\xi^\mu_{(a)}$ are linear combinations of translations and special conformal transformations on the defect plane, while $\xi^\mu_{(\phi)}$ generates rotations in the $(x^1,x^2)$ plane.
In this geometry, there is no issue with the boundary condition at infinity and, consequently,  the expectation values of the $SL(2,\mathbb{R})$ charges on a surface wrapping the defect  (see fig. ~\ref{TorusFig}) 
vanish,
\begin{equation}\label{eq_Qzero}
Q_\xi(D)=0\qquad \text{(circular defect)}\,~.
\end{equation}
The statement~\eqref{eq_Qzero} can be checked using the explcit form of the stress-tensor one-point function in a circular geometry, which depends on a unique constant\footnote{The coefficient $h_D$ is a physical characteristic of the DCFT and it was computed in various supersymmetric examples \cite{Correa:2012at,Fucito:2015ofa,Fiol:2015spa,Bianchi:2018zpb,Bianchi:2019dlw} (for a general approach to supersymmetric line defects see~\cite{Agmon:2020pde}).} $h_D$ (see e.g. \cite{Gomis:2008qa,Billo:2016cpy} for the explicit expressions).
In particular, in contrast to the infinite line,  every point of the surface can be brought arbitrarily close to the defect compatibly with the identity \eqref{eq_Qzero}.\footnote{In spite of this, such a configuration is conformally equivalent to a straight line surrounded by a surface whose radius becomes increasingly large as the line extends to infinity.} For this reason in the following we will focus on circular defects.

\paragraph*{Defect RG} The main goal of this work is to study defect renormalization group (DRG) flows.
A DRG may be triggered perturbing a DCFT with one or more relevant defect operators. For instance,  we may consider a defect operator $\mO$ with $\Delta_{\mO}<1$:
\begin{equation}\label{eq_DCFT_perturbation}
S_{DCFT}\rightarrow S_{DCFT}+ M_0^{1-\Delta_{\mO}}\int_Dd\sigma\mO(\sigma)\,,
\end{equation}
where $\int_D$ stand for integration along the defect and $M_0$ is the mass scale of the flow. Conformal invariance (i.e. $SL(2,\mathbb{R})$ transformations that preserve the defect) is now explicitly broken by the scale $M_0$ to just translations along the defect.

Due to the locality of the bulk CFT, the bulk stress tensor remains conserved and traceless (up to possible bulk trace anomalies in curved space) away from the line. However, now a defect stress tensor $T_D$ is allowed. In other words, energy can now be stored on the defect. Not only $T_D$ is allowed, such an operator must always exist away from the fixed points of the defect. The existence of the operator $T_D$ is the reason that $SL(2,\mathbb{R})$ charges are no longer conserved. Since $T_D$ is localized to the defect, what we mean by saying that $SL(2,\mathbb{R})$ charges are no longer conserved is that, if the charges are integrated on surfaces that intersect the defect, then they are not invariant under small deformations.

Invariance under translations along the defect implies that eq. \eqref{eq_Conservation_DCFT} in the presence of $T_D$ is modified to:
\begin{equation}\label{eq_Conservation_RG}
\nabla_{\mu}  T^{\mu\nu}_b=-\delta_D^{d-1}\dot{X}^\nu \dot{T}_D -\delta_D^{d-1}n^\nu_i  D^i \,,
\end{equation}
where $X^\mu(\sigma)$ is the embedding function describing the defect location and the dot stands for derivatives with respect to the line coordinate $\sigma$, so that $\dot{X}^\nu$ is a tangent vector to the defect.\footnote{Here we are assuming that the defect has a trivial induced submanifold metric $g_D=\dot{X}^\mu\dot{X}^\nu g_{\mu\nu}=1$ to simplify the notation.}  Equation \eqref{eq_Conservation_RG} merely expresses the energy balance between the bulk and the defect.

\paragraph*{Spurion analysis and the dilaton} As it often happens in the study of RG flows, it is useful to promote the renormalization group scale to a function of position $M(\sigma)=M_0 e^{\Phi(\sigma)}$ \cite{Komargodski:2011xv,Luty:2012ww}, where $\Phi(\sigma)$ is a dimensionless background dilaton field. To linear order, the partition function of the theory depends on the dilaton through to the defect energy-momentum tensor:\footnote{In general the conformal symmetry of the line defect is violated by a non-vanishing $T_D$. Such a non-vanishing $T_D$ can be split (not unambiguously) into a c-number due to conformal anomalies and a nontrival local operator. A somewhat common convention, which we also adopt here, is to have the dilaton couplings compensate for the operatorial violation of scale invariance, but not the trace anomalies, which might generically be present in the bulk - see e.g. \cite{Komargodski:2011xv,Jensen:2015swa} for details. }
\begin{align}\nn
&\log Z\vert_{\Phi+\delta\Phi}=\log Z\vert_{\Phi}+\int _D d\sigma\delta\Phi(\sigma)\langle T_D(\sigma)\rangle_{\Phi}+\\ \nn
&+\frac12\int_D d\sigma_1\int_Dd\sigma_2 \delta\Phi(\sigma_1)\delta\Phi(\sigma_2)\langle T_D(\sigma_1)T_D(\sigma_2)\rangle_{\Phi}\\
&
+\ldots\,. \label{eq_Expansion_Z}
\end{align}
The background dilaton field acts a source for the theory. This in turn modifies the conservation equation \eqref{eq_Conservation_RG} as follows \cite{Osborn:1993cr,Cuomo:2021cnb}
\begin{equation}\label{eq_Conservation_Dilaton}
\nabla_{\mu}  T^{\mu\nu}_b =-\delta_D^{d-1}\dot{X}^\nu\left( \dot{T}_D
-\dot{\Phi}\,T_D \right)-\delta_D^{d-1}n^\nu_i D^i \,.
\end{equation}
If one views the coordinate along the defect as time, then a nontrivial $\Phi(\sigma)$ renders the theory time dependent and~\eqref{eq_Conservation_Dilaton} relates the non-conservation rate of the charge associated with translations along the defect with the derivative of the dilaton source.  

A position dependent mass scale breaks the $SL(2,\mathbb{R})$ symmetry completely. What we gain by introducing the general background field $\Phi(\sigma)$ is that $SL(2,\mathbb{R})$ allows us to relate  different theories instead of directly placing constraints on a given theory. Indeed, we will use eq. \eqref{eq_Conservation_Dilaton} in what follows to derive some non-trivial identities relating theories with different values for the source $\Phi(\sigma)$.

\paragraph*{RG flows induced by the broken charges} It is crucial to realize that the identity \eqref{eq_Qzero} holds irrespectively of the breaking of scale invariance on the defect (i.e. it holds for any $\Phi(\sigma)$). This is because the charges wrapping the defect do not intersect it and hence such charges are oblivious to what happens on the defect and they remain invariant under small deformations. They can be moved off to infinity where they annihilate the vacuum. (To see that, one can realize the wrapping surface as the difference between two $S^{d-1}$ surfaces outside and inside the loop.)

As we explained, on general grounds one expects that $SL(2,\mathbb{R})$ transformation can be reabsorbed into a transformation of the dilaton $\Phi(\sigma)$, leading to relations between different theories. 
This can be made precise by using eq. \eqref{eq_Qzero}. 
To that end, consider shrinking the radius of the topological surface enclosing the defect (see fig.~\ref{TorusFig}). 
It is clear from Gauss's law that the only contribution in the integration of the stress tensor arises from the right hand side in eq. \eqref{eq_Conservation_Dilaton}.
We therefore conclude that eq. \eqref{eq_Qzero} implies the following relation:\footnote{We use conventions such that the bulk stress tensor $T^{\mu\nu}_b$ does not contain any contribution proportional to $\delta^{d-1}_D$ and therefore it is traceless everywhere \cite{Jensen:2015swa,Cuomo:2021cnb}.}
\begin{equation}\label{eq_Ward_Pre}
\begin{split}
0&=Q_{\xi}(D)=\int d^{d-1}\Sigma^\mu \langle T_{\mu\nu}^b\rangle\xi^\nu\\
&=\int_D d\sigma\left(\dot{\xi}_D+\xi_D\dot{\Phi}\right)\langle T_D\rangle\,,
\end{split}
\end{equation}
where in the second line we integrated by parts and we denoted by $\xi_D$ the projection of the Killing vectors \eqref{eq_Killing} on the defect.  We crucially used  the fact that the normal components of the $SL(2,\mathbb{R})$ Killing vectors vanish on the defect. In fact, for more general conformal Killing vectors which do not leave the loop invariant, an analogous identity picks an additional contribution from the displacement operator in eq. \eqref{eq_Conservation_Dilaton} but we do not study these identities here. 

Due to the linear coupling between the defect stress tensor and the dilaton, we may interpret eq. \eqref{eq_Ward_Pre} as an equivalence between defects with different DRG scales $M(\sigma)$:
\begin{equation}\label{eq_Equivalence_RGs}
\Phi\sim \Phi+\alpha\left(\dot{\xi}_D+\xi_D\dot{\Phi}\right)\qquad
|\alpha|\ll 1\,,
\end{equation}
for any $SL(2,\mathbb{R})$ Killing vector $\xi$ and any infinitesimal $\alpha$.  This observation is most useful when considering the expansion of the partition function \eqref{eq_Expansion_Z} around $\Phi=0$.
Demanding the equivalence \eqref{eq_Equivalence_RGs} at each order in the field expansion we then find an infinite number of identities for the correlation functions of the defect stress tensor. At second order in the field expansion we obtain the following one (omitting the subscript $\Phi=0$ from now on): 
\begin{align}
&\int_D d\sigma\,\xi_D(\sigma)\dot{\Phi}(\sigma)\langle T_D(\sigma)\rangle =\nn\\
&-\int_D d\sigma_1\int_Dd\sigma_2\, \dot{\xi}_D(\sigma_1)\Phi(\sigma_2)\langle T_D(\sigma_1)T_D(\sigma_2)\rangle\,.
\label{eq_id2}
\end{align}
Crucially, this identity holds for any $\Phi(\sigma)$.  Notice that the right hand side of eq. \eqref{eq_id2} for generic choices of the dilaton profile is naively divergent. Our arguments however ensure that these identities must hold in any regularization scheme which preserves the invariance of the partition function under diffeomorphisms and defect reparametrizations.

At this point it is useful to specify a cylindrical system of coordinates on the defect: $x^1=R\cos\phi$, $x^2=R\sin\phi$ and set $\sigma=R\phi$. The projection of the three Killing vectors in eq. \eqref{eq_Killing} reads, respectively,
\begin{equation}
\xi_D=-\sin\phi~,\quad \xi_D=\cos\phi~,\quad \xi_D=-1~.
\end{equation}

Eq. \eqref{eq_id2} is trivial for $\xi_D=-1$, but provides non-trivial constraints for the other two choices, which lead to identical constraints.  A particularly useful relation is obtained choosing $\xi_D=-\sin\phi$ and $\Phi\propto\cos\phi$ in eq. \eqref{eq_id2}. This leads to:
\begin{multline}\label{eq_Master_Identity}
R\int_Dd\phi\langle T_D(\phi)\rangle\\=R^2
\int_D d\phi_1\int_Dd\phi_2\langle
T_D(\phi_1)T_D(\phi_2)\rangle\cos(\phi_1-\phi_2)\,,
\end{multline}
where we used trigonometric identities and invariance under translations along the defect to simplify both sides.  Eq. \eqref{eq_Master_Identity} will be very useful in providing a gradient formula for the DRG flow of a suitably defined defect entropy. 

\paragraph*{The defect entropy} Our discussion thus far focused on defects in flat space, but all our considerations apply on all conformally equivalent manifolds.  These include the $d$-dimensional sphere of radius $R$, with the defect spanning a maximal circle, and the cylinder $\mathds{R}\times S^{d-1}$, with the defect on the equator of $S^{d-1}$ at a fixed value of the Euclidean time $\tau=\log x^2/R=0$. 

We can use any of these geometries to define a defect $g$-function, $g(M_0 R)$, in terms of the partition function in the presence of the defect, normalized by the partition function without it:
\begin{equation}\label{eq_def_g}
\log g(M_0 R)=
\log Z_{\mathcal{M}}-
\log Z^{(CFT)}_{\mathcal{M}}\,,
\end{equation}
where $\log Z^{(CFT)}_{\mathcal{M}}$ is the partition function of the theory without the defect.\footnote{$\log Z^{(CFT)}_{\mathcal{M}}$ is divergent on the infinite cylinder or in flat space, moreover, in general, it is ambiguous due to various counterterms. But these bulk issues cancel from the definition of $g(M_0 R)$. } 
The defect contribution $g$ depends only on the dimensionless product $M_0 R$ and it reduces to a constant at the fixed points (in a sense that we will explain below).   

We must now ask to what extent is $g$ well defined at the fixed points and away from them. 
$\log g$ can be shifted  by the addition of a cosmological constant counterterm $\int d\sigma M_0\sim M_0 R$ with an arbitrary coefficient.  All other nontrivial geometric invariants which are analytic around the flat metric have dimension larger than one and cannot appear as counterterms. Therefore no additional ambiguities exist in $d>2$ (we will discuss $d=2$ more in detail below). Therefore, one can  obtain a scheme-indepdent quantity which we will refer to as the \emph{defect entropy}, defined as:\footnote{This terminology differs from that of \cite{Kobayashi:2018lil}, where the term \emph{defect entropy} referred to the defect contribution to the entanglement entropy. Note that in $d=2$ the defect entanglement entropy and ordinary entropy coincide at fixed points because $h_D=0$. Here we see that the correct generalization to higher dimensions involves the defect entropy and not the defect entanglement entropy which is also sensitive to $h_D$ \cite{Lewkowycz:2013laa}. }
\begin{equation}\label{eq_Defect_Entropy}
s(M_0 R)=\left(1-R\frac{\partial}{\partial R}\right)\log g(M_0 R)\,.
\end{equation}
At the fixed points, $s(M_0 R)$ is a pure number which is scheme independent. It is equal to the perimeter-independent contribution to $\log g(M_0R)$ at the fixed point. We will refer to these fixed point values of $g$ as $g_{UV}, g_{IR}$, respectively. We will show that $s(M_0 R)$ decreases monotonically under DRG, implying $g_{UV}>g_{IR}$.

In $d=2$ eq. \eqref{eq_Defect_Entropy} coincides with the interface contribution to the thermal entropy of the theory. To make the connection with $d=2$ precise, one needs to remember that in $d=2$ we can also allow the counterterm $\int d\sigma K$, where $K$ is the extrinsic curvature.\footnote{For $d>2$ the extrinsic curvature is not analytic at zero and therefore is not an allowed counterterm; see e.g. \cite{Jensen:2015swa} for a concise review of sub-manifold geometry. The counterterm  $\int d\sigma K$ was discussed in a different context when studying the Entanglement Entropy in 2+1 dimensions, see e.g.~\cite{Grover:2011fa,Liu:2012eea}. } Such a term vanishes for a maximal circle in $S^2$ and on $\mathbb{R}\times S^1$ and therefore all our conclusions hold unaltered on those manifolds.  Furthermore $CPT$ invariance implies that the coefficient of this counterterm should be purely imaginary. Therefore the definition in eq. \eqref{eq_Defect_Entropy} is meaningful also in flat space provided we focus on the real part of the defect entropy.

\paragraph*{The gradient formula} We now have all the ingredients to derive a gradient formula for the DRG flow of the defect entropy.  Since $g$ depends on $M_0R$ only, for constant dilaton $\Phi$,  it follows that $g$ depends on the combination $R M_0 e^\Phi$. We may therefore write the variation of the defect entropy $s$ under a change in the mass scale as follows:
\begin{equation}\label{eq_dS1}
M_0\frac{\partial }{\partial M_0}s(M_0 R)=
\left[
\left(\frac{d}{d\Phi}-\frac{d^2}{d\Phi^2}\right)\log g\left(R M_0e^\Phi\right)\right]_{\Phi=0}\,.
\end{equation}
Using the expansion \eqref{eq_Expansion_Z} for constant $\Phi$ we then can write eq. \eqref{eq_dS1} in terms of correlation functions of the defect stress tensor
\begin{equation}\label{eq_dS2}
\begin{split}
M_0\frac{\partial }{\partial M_0}s(M_0 R)&=R \int_D d\phi\,\langle T_D(\phi)\rangle\\
&-R^2\int_D d\phi_1\int_Dd\phi_2\langle T_D(\phi_1)T_D(\phi_2)\rangle\,.
\end{split}
\end{equation}
Eq. \eqref{eq_dS2} may not seem very useful at first sight. It is not manifestly sign-definite, nor is it manifestly finite. To clarify these issues, we can rewrite the first term using eq. \eqref{eq_Master_Identity}. We obtain:
\begin{equation}\label{eq_gradient}
\begin{aligned}
&M_0\frac{\pd s}{\pd M_0}=\\
&-R^2\int_D d\phi_1 \int_D d\phi_2\langle T_D(\phi_1)T_D(\phi_2)\rangle\left[1-\cos\left(\phi_1-\phi_2\right)\right]\,.
\end{aligned}
\end{equation}
The right hand side of~\eqref{eq_gradient} is free of divergences and ambiguities due to the double zero of $1-\cos\left(\phi_1-\phi_2\right)$.  Furthermore, \eqref{eq_gradient} is manifestly negative in a reflection positive theory (note that this also applies to a connected 2-point function, as on the right hand side of  \eqref{eq_gradient}). Therefore, we deduce that $s$ monotonically decreases along defect RG flows, implying that the UV and IR DCFT satisfy
\begin{equation}\label{eq_g_monotonic}
g_{UV}> g_{IR}\,.
\end{equation}
Eq. \eqref{eq_gradient} additionally implies that $s$ does does no depend on the marginal parameters on the defect.\footnote{It is often the case that $g$ and $s$ depend on the marginal couplings of the bulk CFT \cite{Elitzur:1998va,Fredenhagen:2006dn,Elitzur:2012wm,Bianchi:2019umv,Herzog:2019rke}. Furthermore $g$ and $s$ do not have any obvious monotonicity property under bulk RG flows \cite{Green:2007wr}.}

In $d=2$, equation~\eqref{eq_g_monotonic} was originally conjectured to hold for boundaries (and therefore, using the folding trick, for interfaces) by Affleck and Ludwig \cite{Affleck:1991tk,Affleck:1992ng}. In $d=2$, in the regime where the DRG flow can be described in terms of finitely many couplings and beta functions, a gradient formula equivalent to eq. \eqref{eq_gradient} was proposed in the context of string field theory \cite{Witten:1992qy, Witten:1992cr,Shatashvili:1993kk, Shatashvili:1993ps, Kutasov:2000qp}. It was then  established by Friedan and Konechny \cite{Friedan:2003yc}. An alternative proof of eq. \eqref{eq_g_monotonic} in $d=2$ was also given \cite{Casini:2016fgb} using quantum information methods.\footnote{See, for instance,  also \cite{Yamaguchi:2002pa,Takayanagi:2011zk,Erdmenger:2013dpa} for a holographic setup.} Our work provides an extension of those results to line defects in an arbitrary number of dimensions. We also remark that the inequality \eqref{eq_g_monotonic} was recently conjectured in \cite{Kobayashi:2018lil} for arbitrary $d$.
Another remark is that the trivial line has $g=1$. However, it may a priori be that $g<1$ for some non-trivial lines, as sometimes happens in $2d$ \cite{Oshikawa:1996dj,Oshikawa:1996ww}.

Eq. \eqref{eq_g_monotonic} was extensively checked in $d=2$, see e.g. \cite{Affleck:1991tk,Affleck:1992ng,Affleck:1995ge,Konechny:2003yy}. 
We additionally verified our results \eqref{eq_gradient} and \eqref{eq_g_monotonic} in sevaral concrete examples, including a flow between Wilson lines in $\mathcal{N}=4$ SYM previously studied in \cite{Polchinski:2011im,Beccaria:2017rbe}.  Details can be found in the Supplemental Material. 

Finally, we remark that the partition function of higher-dimensional defects is subject to further ambiguities besides a cosmological constant, rendering a generalization of our arguments not straightforward. For two- and four-dimensional defects irreversibility of the DRG flow was proven via different means, using Weyl anomaly matching \cite{Jensen:2015swa,Wang:2021mdq}.

\begin{acknowledgments}
\paragraph*{Acknowledgements}
We acknowledge useful discussions with Bartolomeu Fiol, Sergei Gukov, Simeon Hellerman, M\'ark Mezei, Luigi Tizzano,  Cumrun Vafa, and  Yifan Wang. GC is supported by the Simons Foundation (Simons Collaboration on the Non-perturbative Bootstrap) grants 488647 and 397411. ZK and ARM are supported in part by the Simons Foundation grant 488657 (Simons Collaboration on the Non-Perturbative Bootstrap) and the BSF grant no. 2018204. The work of ARM was also supported in part by the Zuckerman-CHE STEM Leadership Program.
\end{acknowledgments}

\appendix

\section{Supplemental Material}

\subsection{A. Subtleties for the infinite line geometry}

Naively, a defect on an infinite straight line is conformally equivalent to a circular defect.  However, there is a subtlety associated to the infinite extent of the straight line defect. Here we detail this issue.

To explain the subtlety with straight line defects, consider a defect $D$ which extends in the $d$th direction at $x^i=0$, where $i,j,\ldots$ denote indices $1,\ldots,d-1$ transverse to the line.  For $d>2$, the one-point function of the stress tensor  depends on a constant $h_D$ and reads \cite{Kapustin:2005py}:
\begin{equation}\label{eq_Stress_Tensor_1pt}
\begin{gathered}
\langle T^{dd}(x)\rangle= h_D\frac{d-2}{r^d}\,,\quad
\langle T^{id}(x)\rangle=0\,,\\
\langle T^{ij}(x)\rangle=- h_D\frac{(2\,\delta^{ij}- d\, x^i x^j/r^2)}{r^d}\,,
\end{gathered}
\end{equation}
where $r^2=x^ix^i$ is the distance from the line operator (we assume that the defect has zero transverse spin for simplicity).

Eq. \eqref{eq_Stress_Tensor_1pt} is covariant under conformal transformations. However, a subtlety arises when we consider the expectation values of the $SL(2,\mathbb{R})$ charges. The expectation values of the charges are obtained by integrating the stress tensor contracted with the appropriate Killing vector at a fixed distance $\varepsilon$ on a cylinder around the defect:
\begin{equation}\label{eq_chargeS}
Q_\xi(D)=\int_{r=\varepsilon} d^{d-1}\Sigma^\mu\langle T^b_{\mu\nu}\rangle\xi^\nu\,.
\end{equation}
For the dilation and special conformal Killing vectors along the line one finds a result proportional to $1/\varepsilon$ times a linearly divergent integral $\int d x^d$. This is in sharp contrast with the expected charge conservation.  Technically, this is because the $x\rightarrow\infty $ limit of the stress tensor depends on the distance $r$ from the defect,  and it is therefore not single-valued at the point at infinity.  This implies that that the conformal charges vanish only when the distance between the integration surface and the defect diverges sufficiently fast as $x^d\rightarrow\pm \infty$. This problem with the conformal charges is presumably related to the disagreement between the expectation value of circular and linear Maldacena-Wilson loops in $\mathcal{N}=4$ SYM \cite{Erickson:2000af, Drukker:2000rr,Pestun:2007rz}. 

\subsection{B. Examples}

The inequality $g_{UV}>g_{IR}$ was extensively checked in the literature in $d=2$. Early examples include flows between the free and the fixed boundary conditions in the Ising model \cite{Affleck:1991tk} and the weak to strong coupling flow in the Kondo effect \cite{Affleck:1992ng,Affleck:1995ge}. In $d>2$ flows between various Wilson lines in 4 dimensions and a vast class of holographic flows was considered in \cite{Kobayashi:2018lil}. The results are consistent with the monotonicity of $g$.

Let us now discuss the gradient formula:
\begin{equation}\label{eq_gradient2}
\begin{aligned}
&M_0\frac{\pd s}{\pd M_0}=\\
&-R^2\int_D d\phi_1 \int_D d\phi_2\langle T_D(\phi_1)T_D(\phi_2)\rangle\left[1-\cos\left(\phi_1-\phi_2\right)\right]\,,
\end{aligned}
\end{equation}
where $s$ is the defect entropy defined in the main text.
We verified eq. \eqref{eq_gradient2} in two concrete examples: conformal perturbation theory and the flow from a standard Wilson loop to a supersymmetric one in $\mathcal{N}=4$ SYM, proposed by Polchinski and Sully \cite{Polchinski:2011im}, at weak 't Hooft coupling.

In conformal perturbation theory, one starts from an abstract DCFT and perturbs it with one or more weakly or marginally relevant operators; one then computes the partition function expanding in the couplings of the perturbations. It is then simple to compute the defect entropy and verify the gradient formula \eqref{eq_gradient2} using $T_D=\beta_i\mO_i$, where $\beta_i$ are the beta functions of the defect coupling and $\mO_i$ the defect operators with which the DCFT is deformed.
In this setup, in $d=2$, eq. \eqref{eq_gradient2} was verified to third order in perturbation theory in \cite{Konechny:2003yy}.
A similar argument holds for any $d$.

Let us now consider the DRG flow from the ordinary Wilson loop (WL) to the $1/2$ BPS Wilson-Maldacena loop (WML) in $\mathcal{N}=4$ SYM in four dimensions in the planar limit.  This was proposed in \cite{Polchinski:2011im} and studied in detail in \cite{Beccaria:2017rbe}. One considers a single-parameter family of Wilson loop operators in the fundamental representation
\begin{equation}\label{eq_Wilson_Loop_N4}
W^{(\zeta)} = \frac{1}{N}\text{Tr}\, \mathcal{P}\, \text{exp}\, \oint_C d\tau \left[iA_\mu(x)\dot{x}^\mu+ \zeta\Phi_m(x)\theta^m|\dot{x}| \right],
\end{equation}
where $\theta_m^2=1$, $\zeta$ is the coefficient in front of the scalar coupling ($m=1,\cdots,6$) and we follow closely the notations of \cite{Beccaria:2017rbe},  focusing on circular contours.  The theory admits a UV fixed point $\zeta=0$ and may flow to the WML ($|\zeta|=1$), that provides an IR stable fixed point. 
Generically, the expectation values of
the Wilson loop operators depend on the renormalization scale through:
\begin{equation}
\begin{aligned}
&\langle W^{(\zeta)} \rangle  \equiv W\left(\lambda; \zeta(M_0R),M_0R\right), \\
& M_0\frac{\partial}{\partial M_0}W+ \beta_\zeta \frac{\partial}{\partial\zeta}W=0,
\end{aligned}
\end{equation}
where we note that the boundary coupling $\zeta$  depends on the renormalization scale $\zeta =\zeta(M_0R) $. 
Here $\langle W^{(\zeta)}\rangle$ stands for the $g$-function of the defect.
At weak 't Hooft coupling, $\lambda \ll 1$, this dependence follows from the beta function of $\zeta$ \cite{Polchinski:2011im}:
\begin{equation}
\label{eq:betaFunctionWeakCoup}
\beta_\zeta =M_0\frac{\partial \zeta}{\partial M_0} 
=-\frac{\lambda}{8\pi^2}\zeta\left( 1-\zeta^2\right)+\mathcal{O}\left(\lambda^2\right)\,.
\end{equation}  
The partition function of the DCFT was evaluated to order $\mO(\lambda^2)$ in \cite{Beccaria:2017rbe}, where it was found that (in units such that $R=1$):
\begin{equation}\label{eq:gWeakCoup}
\langle W^{(\zeta)} \rangle = 1+ \frac{1}{8}\lambda+\left[\frac{1}{192}+\frac{1}{128\pi^2}\left(1-\zeta^2 \right)^2 \right]\lambda^2 +\mO(\lambda^3)\,.
\end{equation}
Notice that these results, though perturbative in $\lambda$, are exact throughout the flow. From eq. \eqref{eq:gWeakCoup} one clearly reads $g_{UV}>g_{IR}$.  Upon taking derivatives with respect to $M_0$ and using eq. \eqref{eq:betaFunctionWeakCoup}, one may then compute the DRG gradient of the defect entropy to be:
\footnote{For that we use: $s(M_0 R)=\left( 1-R\frac{\partial}{\partial R}\right)g(M_0R)=\log\langle W^{(\zeta)} \rangle+\beta_\zeta \frac{\partial }{\partial\zeta}\log \langle W^{(\zeta)} \rangle$, 
and $M_0\frac{\partial s}{\partial M_0}=-\beta_\zeta\frac{\partial s}{\partial \zeta}$, where $\beta_\zeta$ is given by \eqref{eq:betaFunctionWeakCoup}.}
\begin{equation}
\label{eq:GradientForWLRG}
M_0 \frac{\partial s}{\partial M_0} = -\frac{\lambda^3}{256\pi^4}\zeta^2\left( 1-\zeta^2\right)^2\,+\mO\left(\lambda^4\right).
\end{equation}
Hence $s$ monotonically decreases along the DRG flow.  Furthermore, since to the order we are working we may neglect anomalous dimensions,  we easily find the two-point function of the defect stress tensor \cite{Beccaria:2017rbe}:
\begin{equation}
\langle T_D(\phi)T_D(0)\rangle=\frac{\lambda}{8\pi^2}\frac{\beta_{\zeta}^2}{\left(2\sin\frac{\phi}{2}\right)^2}\left[1+\mO\left(\lambda\right)\right]\,.
\end{equation}
It is then simple to use this equation and the beta-function \eqref{eq:betaFunctionWeakCoup} to verify explicitly the agreement of the RHS of the gradient formula \eqref{eq_gradient2} with the expression \eqref{eq:GradientForWLRG}. This provides a test of the gradient formula \eqref{eq_gradient2} in the small 't Hooft coupling regime.

In the strong coupling regime one expects a similar
DRG flow to take place. This is supported by the holographic calculation of the DCFT partition functions at the fixed points, whose ratio satisfies $\langle W^{(0)} \rangle/\langle W^{(1)} \rangle\sim\lambda^2$ for $\lambda\gg 1$ \cite{Beccaria:2017rbe}.

\bibliography{Biblio}

\providecommand{\href}[2]{#2}\begingroup\raggedright\begin{thebibliography}{10}

\bibitem{1986JETPL..43..730Z}
A.~B. {Zomolodchikov}, \emph{{``Irreversibility'' of the flux of the
  renormalization group in a 2D field theory}}, {\emph{Soviet Journal of
  Experimental and Theoretical Physics Letters} {\bfseries 43} (1986) 730}.

\bibitem{Cardy:1988cwa}
J.~L. Cardy, \emph{{Is There a c Theorem in Four-Dimensions?}},
  \href{https://doi.org/10.1016/0370-2693(88)90054-8}{\emph{Phys. Lett. B}
  {\bfseries 215} (1988) 749}.

\bibitem{Cappelli:1990yc}
A.~Cappelli, D.~Friedan and J.~I. Latorre, \emph{{C theorem and spectral
  representation}},
  \href{https://doi.org/10.1016/0550-3213(91)90102-4}{\emph{Nucl. Phys. B}
  {\bfseries 352} (1991) 616}.

\bibitem{Osborn:1991gm}
H.~Osborn, \emph{{Weyl consistency conditions and a local renormalization group
  equation for general renormalizable field theories}},
  \href{https://doi.org/10.1016/0550-3213(91)80030-P}{\emph{Nucl. Phys. B}
  {\bfseries 363} (1991) 486}.

\bibitem{Myers:2010tj}
R.~C. Myers and A.~Sinha, \emph{{Holographic c-theorems in arbitrary
  dimensions}}, \href{https://doi.org/10.1007/JHEP01(2011)125}{\emph{JHEP}
  {\bfseries 01} (2011) 125} [\href{https://arxiv.org/abs/1011.5819}{{\ttfamily
  1011.5819}}].

\bibitem{Jafferis:2011zi}
D.~L. Jafferis, I.~R. Klebanov, S.~S. Pufu and B.~R. Safdi, \emph{{Towards the
  F-Theorem: N=2 Field Theories on the Three-Sphere}},
  \href{https://doi.org/10.1007/JHEP06(2011)102}{\emph{JHEP} {\bfseries 06}
  (2011) 102} [\href{https://arxiv.org/abs/1103.1181}{{\ttfamily 1103.1181}}].

\bibitem{Komargodski:2011vj}
Z.~Komargodski and A.~Schwimmer, \emph{{On Renormalization Group Flows in Four
  Dimensions}}, \href{https://doi.org/10.1007/JHEP12(2011)099}{\emph{JHEP}
  {\bfseries 12} (2011) 099} [\href{https://arxiv.org/abs/1107.3987}{{\ttfamily
  1107.3987}}].

\bibitem{Casini:2012ei}
H.~Casini and M.~Huerta, \emph{{On the RG running of the entanglement entropy
  of a circle}}, \href{https://doi.org/10.1103/PhysRevD.85.125016}{\emph{Phys.
  Rev. D} {\bfseries 85} (2012) 125016}
  [\href{https://arxiv.org/abs/1202.5650}{{\ttfamily 1202.5650}}].

\bibitem{Elvang:2012st}
H.~Elvang, D.~Z. Freedman, L.-Y. Hung, M.~Kiermaier, R.~C. Myers and
  S.~Theisen, \emph{{On renormalization group flows and the a-theorem in 6d}},
  \href{https://doi.org/10.1007/JHEP10(2012)011}{\emph{JHEP} {\bfseries 10}
  (2012) 011} [\href{https://arxiv.org/abs/1205.3994}{{\ttfamily 1205.3994}}].

\bibitem{Elvang:2012yc}
H.~Elvang and T.~M. Olson, \emph{{RG flows in d dimensions, the dilaton
  effective action, and the a-theorem}},
  \href{https://doi.org/10.1007/JHEP03(2013)034}{\emph{JHEP} {\bfseries 03}
  (2013) 034} [\href{https://arxiv.org/abs/1209.3424}{{\ttfamily 1209.3424}}].

\bibitem{Yonekura:2012kb}
K.~Yonekura, \emph{{Perturbative c-theorem in d-dimensions}},
  \href{https://doi.org/10.1007/JHEP04(2013)011}{\emph{JHEP} {\bfseries 04}
  (2013) 011} [\href{https://arxiv.org/abs/1212.3028}{{\ttfamily 1212.3028}}].

\bibitem{Antipin:2013pya}
O.~Antipin, M.~Gillioz, E.~M\o{}lgaard and F.~Sannino, \emph{{The a theorem for
  gauge-Yukawa theories beyond Banks-Zaks fixed point}},
  \href{https://doi.org/10.1103/PhysRevD.87.125017}{\emph{Phys. Rev. D}
  {\bfseries 87} (2013) 125017}
  [\href{https://arxiv.org/abs/1303.1525}{{\ttfamily 1303.1525}}].

\bibitem{Grinstein:2013cka}
B.~Grinstein, A.~Stergiou and D.~Stone, \emph{{Consequences of Weyl Consistency
  Conditions}}, \href{https://doi.org/10.1007/JHEP11(2013)195}{\emph{JHEP}
  {\bfseries 11} (2013) 195} [\href{https://arxiv.org/abs/1308.1096}{{\ttfamily
  1308.1096}}].

\bibitem{Jack:2013sha}
I.~Jack and H.~Osborn, \emph{{Constraints on RG Flow for Four Dimensional
  Quantum Field Theories}},
  \href{https://doi.org/10.1016/j.nuclphysb.2014.03.018}{\emph{Nucl. Phys. B}
  {\bfseries 883} (2014) 425}
  [\href{https://arxiv.org/abs/1312.0428}{{\ttfamily 1312.0428}}].

\bibitem{Baume:2014rla}
F.~Baume, B.~Keren-Zur, R.~Rattazzi and L.~Vitale, \emph{{The local
  Callan-Symanzik equation: structure and applications}},
  \href{https://doi.org/10.1007/JHEP08(2014)152}{\emph{JHEP} {\bfseries 08}
  (2014) 152} [\href{https://arxiv.org/abs/1401.5983}{{\ttfamily 1401.5983}}].

\bibitem{Grinstein:2014xba}
B.~Grinstein, D.~Stone, A.~Stergiou and M.~Zhong, \emph{{Challenge to the $a$
  Theorem in Six Dimensions}},
  \href{https://doi.org/10.1103/PhysRevLett.113.231602}{\emph{Phys. Rev. Lett.}
  {\bfseries 113} (2014) 231602}
  [\href{https://arxiv.org/abs/1406.3626}{{\ttfamily 1406.3626}}].

\bibitem{Giombi:2014xxa}
S.~Giombi and I.~R. Klebanov, \emph{{Interpolating between $a$ and $F$}},
  \href{https://doi.org/10.1007/JHEP03(2015)117}{\emph{JHEP} {\bfseries 03}
  (2015) 117} [\href{https://arxiv.org/abs/1409.1937}{{\ttfamily 1409.1937}}].

\bibitem{Jack:2015tka}
I.~Jack, D.~R.~T. Jones and C.~Poole, \emph{{Gradient flows in three
  dimensions}}, \href{https://doi.org/10.1007/JHEP09(2015)061}{\emph{JHEP}
  {\bfseries 09} (2015) 061}
  [\href{https://arxiv.org/abs/1505.05400}{{\ttfamily 1505.05400}}].

\bibitem{Cordova:2015fha}
C.~Cordova, T.~T. Dumitrescu and K.~Intriligator, \emph{{Anomalies,
  renormalization group flows, and the a-theorem in six-dimensional (1, 0)
  theories}}, \href{https://doi.org/10.1007/JHEP10(2016)080}{\emph{JHEP}
  {\bfseries 10} (2016) 080}
  [\href{https://arxiv.org/abs/1506.03807}{{\ttfamily 1506.03807}}].

\bibitem{Casini:2015woa}
H.~Casini, M.~Huerta, R.~C. Myers and A.~Yale, \emph{{Mutual information and
  the F-theorem}}, \href{https://doi.org/10.1007/JHEP10(2015)003}{\emph{JHEP}
  {\bfseries 10} (2015) 003}
  [\href{https://arxiv.org/abs/1506.06195}{{\ttfamily 1506.06195}}].

\bibitem{Pufu:2016zxm}
S.~S. Pufu, \emph{{The F-Theorem and F-Maximization}},
  \href{https://doi.org/10.1088/1751-8121/aa6765}{\emph{J. Phys. A} {\bfseries
  50} (2017) 443008} [\href{https://arxiv.org/abs/1608.02960}{{\ttfamily
  1608.02960}}].

\bibitem{Casini:2017vbe}
H.~Casini, E.~Test\'e and G.~Torroba, \emph{{Markov Property of the Conformal
  Field Theory Vacuum and the a Theorem}},
  \href{https://doi.org/10.1103/PhysRevLett.118.261602}{\emph{Phys. Rev. Lett.}
  {\bfseries 118} (2017) 261602}
  [\href{https://arxiv.org/abs/1704.01870}{{\ttfamily 1704.01870}}].

\bibitem{Fluder:2020pym}
M.~Fluder and C.~F. Uhlemann, \emph{{Evidence for a 5d F-theorem}},
  \href{https://doi.org/10.1007/JHEP02(2021)192}{\emph{JHEP} {\bfseries 02}
  (2021) 192} [\href{https://arxiv.org/abs/2011.00006}{{\ttfamily
  2011.00006}}].

\bibitem{Delacretaz:2021ufg}
L.~V. Delacretaz, A.~L. Fitzpatrick, E.~Katz and M.~T. Walters,
  \emph{{Thermalization and Hydrodynamics of Two-Dimensional Quantum Field
  Theories}},  \href{https://arxiv.org/abs/2105.02229}{{\ttfamily 2105.02229}}.

\bibitem{Kapustin:2005py}
A.~Kapustin, \emph{{Wilson-'t Hooft operators in four-dimensional gauge
  theories and S-duality}},
  \href{https://doi.org/10.1103/PhysRevD.74.025005}{\emph{Phys. Rev. D}
  {\bfseries 74} (2006) 025005}
  [\href{https://arxiv.org/abs/hep-th/0501015}{{\ttfamily hep-th/0501015}}].

\bibitem{Gomis:2006sb}
J.~Gomis and F.~Passerini, \emph{{Holographic Wilson Loops}},
  \href{https://doi.org/10.1088/1126-6708/2006/08/074}{\emph{JHEP} {\bfseries
  08} (2006) 074} [\href{https://arxiv.org/abs/hep-th/0604007}{{\ttfamily
  hep-th/0604007}}].

\bibitem{Billo:2013jda}
M.~Bill\'o, M.~Caselle, D.~Gaiotto, F.~Gliozzi, M.~Meineri and R.~Pellegrini,
  \emph{{Line defects in the 3d Ising model}},
  \href{https://doi.org/10.1007/JHEP07(2013)055}{\emph{JHEP} {\bfseries 07}
  (2013) 055} [\href{https://arxiv.org/abs/1304.4110}{{\ttfamily 1304.4110}}].

\bibitem{Gaiotto:2013nva}
D.~Gaiotto, D.~Mazac and M.~F. Paulos, \emph{{Bootstrapping the 3d Ising twist
  defect}}, \href{https://doi.org/10.1007/JHEP03(2014)100}{\emph{JHEP}
  {\bfseries 03} (2014) 100} [\href{https://arxiv.org/abs/1310.5078}{{\ttfamily
  1310.5078}}].

\bibitem{Giombi:2021uae}
S.~Giombi, E.~Helfenberger, Z.~Ji and H.~Khanchandani, \emph{{Monodromy Defects
  from Hyperbolic Space}},  \href{https://arxiv.org/abs/2102.11815}{{\ttfamily
  2102.11815}}.

\bibitem{liu2021magnetic}
S.~Liu, H.~Shapourian, A.~Vishwanath and M.~A. Metlitski, \emph{{Magnetic
  impurities at quantum critical points: large-$N$ expansion and SPT
  connections}},  \href{https://arxiv.org/abs/2104.15026}{{\ttfamily
  2104.15026}}.

\bibitem{tsvelick1985exact}
A.~Tsvelick and P.~Wiegmann, \emph{Exact solution of the multichannel kondo
  problem, scaling, and integrability}, {\emph{Journal of Statistical Physics}
  {\bfseries 38} (1985) 125}.

\bibitem{Affleck:1992ng}
I.~Affleck and A.~W.~W. Ludwig, \emph{{Exact conformal-field-theory results on
  the multichannel Kondo effect: Single-fermion Green\textquoteright{}s
  function, self-energy, and resistivity}},
  \href{https://doi.org/10.1103/PhysRevB.48.7297}{\emph{Phys. Rev. B}
  {\bfseries 48} (1993) 7297}.

\bibitem{Affleck:1995ge}
I.~Affleck, \emph{{Conformal field theory approach to the Kondo effect}},
  {\emph{Acta Phys. Polon. B} {\bfseries 26} (1995) 1869}
  [\href{https://arxiv.org/abs/cond-mat/9512099}{{\ttfamily
  cond-mat/9512099}}].

\bibitem{sachdev1999quantum}
S.~Sachdev, C.~Buragohain and M.~Vojta, \emph{Quantum impurity in a nearly
  critical two-dimensional antiferromagnet}, {\emph{Science} {\bfseries 286}
  (1999) 2479}.

\bibitem{vojta2000quantum}
M.~Vojta, C.~Buragohain and S.~Sachdev, \emph{Quantum impurity dynamics in
  two-dimensional antiferromagnets and superconductors}, {\emph{Physical Review
  B} {\bfseries 61} (2000) 15152}.

\bibitem{Affleck:1991tk}
I.~Affleck and A.~W.~W. Ludwig, \emph{{Universal noninteger 'ground state
  degeneracy' in critical quantum systems}},
  \href{https://doi.org/10.1103/PhysRevLett.67.161}{\emph{Phys. Rev. Lett.}
  {\bfseries 67} (1991) 161}.

\bibitem{Dorey:1999cj}
P.~Dorey, I.~Runkel, R.~Tateo and G.~Watts, \emph{{g function flow in perturbed
  boundary conformal field theories}},
  \href{https://doi.org/10.1016/S0550-3213(99)00772-5}{\emph{Nucl. Phys. B}
  {\bfseries 578} (2000) 85}
  [\href{https://arxiv.org/abs/hep-th/9909216}{{\ttfamily hep-th/9909216}}].

\bibitem{Yamaguchi:2002pa}
S.~Yamaguchi, \emph{{Holographic RG flow on the defect and g theorem}},
  \href{https://doi.org/10.1088/1126-6708/2002/10/002}{\emph{JHEP} {\bfseries
  10} (2002) 002} [\href{https://arxiv.org/abs/hep-th/0207171}{{\ttfamily
  hep-th/0207171}}].

\bibitem{Friedan:2003yc}
D.~Friedan and A.~Konechny, \emph{{On the boundary entropy of one-dimensional
  quantum systems at low temperature}},
  \href{https://doi.org/10.1103/PhysRevLett.93.030402}{\emph{Phys. Rev. Lett.}
  {\bfseries 93} (2004) 030402}
  [\href{https://arxiv.org/abs/hep-th/0312197}{{\ttfamily hep-th/0312197}}].

\bibitem{Azeyanagi:2007qj}
T.~Azeyanagi, A.~Karch, T.~Takayanagi and E.~G. Thompson, \emph{{Holographic
  calculation of boundary entropy}},
  \href{https://doi.org/10.1088/1126-6708/2008/03/054}{\emph{JHEP} {\bfseries
  03} (2008) 054} [\href{https://arxiv.org/abs/0712.1850}{{\ttfamily
  0712.1850}}].

\bibitem{Takayanagi:2011zk}
T.~Takayanagi, \emph{{Holographic Dual of BCFT}},
  \href{https://doi.org/10.1103/PhysRevLett.107.101602}{\emph{Phys. Rev. Lett.}
  {\bfseries 107} (2011) 101602}
  [\href{https://arxiv.org/abs/1105.5165}{{\ttfamily 1105.5165}}].

\bibitem{Estes:2014hka}
J.~Estes, K.~Jensen, A.~O'Bannon, E.~Tsatis and T.~Wrase, \emph{{On Holographic
  Defect Entropy}}, \href{https://doi.org/10.1007/JHEP05(2014)084}{\emph{JHEP}
  {\bfseries 05} (2014) 084} [\href{https://arxiv.org/abs/1403.6475}{{\ttfamily
  1403.6475}}].

\bibitem{Gaiotto:2014gha}
D.~Gaiotto, \emph{{Boundary F-maximization}},
  \href{https://arxiv.org/abs/1403.8052}{{\ttfamily 1403.8052}}.

\bibitem{Jensen:2015swa}
K.~Jensen and A.~O'Bannon, \emph{{Constraint on Defect and Boundary
  Renormalization Group Flows}},
  \href{https://doi.org/10.1103/PhysRevLett.116.091601}{\emph{Phys. Rev. Lett.}
  {\bfseries 116} (2016) 091601}
  [\href{https://arxiv.org/abs/1509.02160}{{\ttfamily 1509.02160}}].

\bibitem{Casini:2016fgb}
H.~Casini, I.~Salazar~Landea and G.~Torroba, \emph{{The g-theorem and quantum
  information theory}},
  \href{https://doi.org/10.1007/JHEP10(2016)140}{\emph{JHEP} {\bfseries 10}
  (2016) 140} [\href{https://arxiv.org/abs/1607.00390}{{\ttfamily
  1607.00390}}].

\bibitem{Andrei:2018die}
N.~Andrei et~al., \emph{{Boundary and Defect CFT: Open Problems and
  Applications}}, \href{https://doi.org/10.1088/1751-8121/abb0fe}{\emph{J.
  Phys. A} {\bfseries 53} (2020) 453002}
  [\href{https://arxiv.org/abs/1810.05697}{{\ttfamily 1810.05697}}].

\bibitem{Kobayashi:2018lil}
N.~Kobayashi, T.~Nishioka, Y.~Sato and K.~Watanabe, \emph{{Towards a
  $C$-theorem in defect CFT}},
  \href{https://doi.org/10.1007/JHEP01(2019)039}{\emph{JHEP} {\bfseries 01}
  (2019) 039} [\href{https://arxiv.org/abs/1810.06995}{{\ttfamily
  1810.06995}}].

\bibitem{Casini:2018nym}
H.~Casini, I.~Salazar~Landea and G.~Torroba, \emph{{Irreversibility in quantum
  field theories with boundaries}},
  \href{https://doi.org/10.1007/JHEP04(2019)166}{\emph{JHEP} {\bfseries 04}
  (2019) 166} [\href{https://arxiv.org/abs/1812.08183}{{\ttfamily
  1812.08183}}].

\bibitem{Giombi:2020rmc}
S.~Giombi and H.~Khanchandani, \emph{{CFT in AdS and boundary RG flows}},
  \href{https://doi.org/10.1007/JHEP11(2020)118}{\emph{JHEP} {\bfseries 11}
  (2020) 118} [\href{https://arxiv.org/abs/2007.04955}{{\ttfamily
  2007.04955}}].

\bibitem{Wang:2020xkc}
Y.~Wang, \emph{{Surface Defect, Anomalies and $b$-Extremization}},
  \href{https://arxiv.org/abs/2012.06574}{{\ttfamily 2012.06574}}.

\bibitem{Nishioka:2021uef}
T.~Nishioka and Y.~Sato, \emph{{Free energy and defect $C$-theorem in free
  scalar theory}}, \href{https://doi.org/10.1007/JHEP05(2021)074}{\emph{JHEP}
  {\bfseries 05} (2021) 074}
  [\href{https://arxiv.org/abs/2101.02399}{{\ttfamily 2101.02399}}].

\bibitem{Wang:2021mdq}
Y.~Wang, \emph{{Defect $a$-Theorem and $a$-Maximization}},
  \href{https://arxiv.org/abs/2101.12648}{{\ttfamily 2101.12648}}.

\bibitem{Sato:2021eqo}
Y.~Sato, \emph{{Free energy and defect $C$-theorem in free fermion}},
  \href{https://doi.org/10.1007/JHEP05(2021)202}{\emph{JHEP} {\bfseries 05}
  (2021) 202} [\href{https://arxiv.org/abs/2102.11468}{{\ttfamily
  2102.11468}}].

\bibitem{Nakayama:2012ed}
Y.~Nakayama, \emph{{Is boundary conformal in CFT?}},
  \href{https://doi.org/10.1103/PhysRevD.87.046005}{\emph{Phys. Rev. D}
  {\bfseries 87} (2013) 046005}
  [\href{https://arxiv.org/abs/1210.6439}{{\ttfamily 1210.6439}}].

\bibitem{Herzog:2017xha}
C.~P. Herzog and K.-W. Huang, \emph{{Boundary Conformal Field Theory and a
  Boundary Central Charge}},
  \href{https://doi.org/10.1007/JHEP10(2017)189}{\emph{JHEP} {\bfseries 10}
  (2017) 189} [\href{https://arxiv.org/abs/1707.06224}{{\ttfamily
  1707.06224}}].

\bibitem{Cuomo:2021cnb}
G.~Cuomo, M.~Mezei and A.~Raviv-Moshe, \emph{{Boundary Conformal Field Theory
  at Large Charge}},  \href{https://arxiv.org/abs/2108.06579}{{\ttfamily
  2108.06579}}.

\bibitem{Osborn:1993cr}
H.~Osborn and A.~C. Petkou, \emph{{Implications of conformal invariance in
  field theories for general dimensions}},
  \href{https://doi.org/10.1006/aphy.1994.1045}{\emph{Annals Phys.} {\bfseries
  231} (1994) 311} [\href{https://arxiv.org/abs/hep-th/9307010}{{\ttfamily
  hep-th/9307010}}].

\bibitem{Gliozzi:2015qsa}
F.~Gliozzi, P.~Liendo, M.~Meineri and A.~Rago, \emph{{Boundary and Interface
  CFTs from the Conformal Bootstrap}},
  \href{https://doi.org/10.1007/JHEP05(2015)036}{\emph{JHEP} {\bfseries 05}
  (2015) 036} [\href{https://arxiv.org/abs/1502.07217}{{\ttfamily
  1502.07217}}].

\bibitem{Liendo:2012hy}
P.~Liendo, L.~Rastelli and B.~C. van Rees, \emph{{The Bootstrap Program for
  Boundary CFT$_d$}},
  \href{https://doi.org/10.1007/JHEP07(2013)113}{\emph{JHEP} {\bfseries 07}
  (2013) 113} [\href{https://arxiv.org/abs/1210.4258}{{\ttfamily 1210.4258}}].

\bibitem{Billo:2016cpy}
M.~Bill\`o, V.~Gon\c{c}alves, E.~Lauria and M.~Meineri, \emph{{Defects in
  conformal field theory}},
  \href{https://doi.org/10.1007/JHEP04(2016)091}{\emph{JHEP} {\bfseries 04}
  (2016) 091} [\href{https://arxiv.org/abs/1601.02883}{{\ttfamily
  1601.02883}}].

\bibitem{Lauria:2020emq}
E.~Lauria, P.~Liendo, B.~C. Van~Rees and X.~Zhao, \emph{{Line and surface
  defects for the free scalar field}},
  \href{https://doi.org/10.1007/JHEP01(2021)060}{\emph{JHEP} {\bfseries 01}
  (2021) 060} [\href{https://arxiv.org/abs/2005.02413}{{\ttfamily
  2005.02413}}].

\bibitem{Belavin:1984vu}
A.~A. Belavin, A.~M. Polyakov and A.~B. Zamolodchikov, \emph{{Infinite
  Conformal Symmetry in Two-Dimensional Quantum Field Theory}},
  \href{https://doi.org/10.1016/0550-3213(84)90052-X}{\emph{Nucl. Phys. B}
  {\bfseries 241} (1984) 333}.

\bibitem{Rattazzi:2008pe}
R.~Rattazzi, V.~S. Rychkov, E.~Tonni and A.~Vichi, \emph{{Bounding scalar
  operator dimensions in 4D CFT}},
  \href{https://doi.org/10.1088/1126-6708/2008/12/031}{\emph{JHEP} {\bfseries
  12} (2008) 031} [\href{https://arxiv.org/abs/0807.0004}{{\ttfamily
  0807.0004}}].

\bibitem{Poland:2018epd}
D.~Poland, S.~Rychkov and A.~Vichi, \emph{{The Conformal Bootstrap: Theory,
  Numerical Techniques, and Applications}},
  \href{https://doi.org/10.1103/RevModPhys.91.015002}{\emph{Rev. Mod. Phys.}
  {\bfseries 91} (2019) 015002}
  [\href{https://arxiv.org/abs/1805.04405}{{\ttfamily 1805.04405}}].

\bibitem{Correa:2012at}
D.~Correa, J.~Henn, J.~Maldacena and A.~Sever, \emph{{An exact formula for the
  radiation of a moving quark in N=4 super Yang Mills}},
  \href{https://doi.org/10.1007/JHEP06(2012)048}{\emph{JHEP} {\bfseries 06}
  (2012) 048} [\href{https://arxiv.org/abs/1202.4455}{{\ttfamily 1202.4455}}].

\bibitem{Fucito:2015ofa}
F.~Fucito, J.~F. Morales and R.~Poghossian, \emph{{Wilson loops and chiral
  correlators on squashed spheres}},
  \href{https://doi.org/10.1007/JHEP11(2015)064}{\emph{JHEP} {\bfseries 11}
  (2015) 064} [\href{https://arxiv.org/abs/1507.05426}{{\ttfamily
  1507.05426}}].

\bibitem{Fiol:2015spa}
B.~Fiol, E.~Gerchkovitz and Z.~Komargodski, \emph{{Exact Bremsstrahlung
  Function in $N=2$ Superconformal Field Theories}},
  \href{https://doi.org/10.1103/PhysRevLett.116.081601}{\emph{Phys. Rev. Lett.}
  {\bfseries 116} (2016) 081601}
  [\href{https://arxiv.org/abs/1510.01332}{{\ttfamily 1510.01332}}].

\bibitem{Bianchi:2018zpb}
L.~Bianchi, M.~Lemos and M.~Meineri, \emph{{Line Defects and Radiation in
  $\mathcal{N}=2$ Conformal Theories}},
  \href{https://doi.org/10.1103/PhysRevLett.121.141601}{\emph{Phys. Rev. Lett.}
  {\bfseries 121} (2018) 141601}
  [\href{https://arxiv.org/abs/1805.04111}{{\ttfamily 1805.04111}}].

\bibitem{Bianchi:2019dlw}
L.~Bianchi, M.~Bill\`o, F.~Galvagno and A.~Lerda, \emph{{Emitted Radiation and
  Geometry}}, \href{https://doi.org/10.1007/JHEP01(2020)075}{\emph{JHEP}
  {\bfseries 01} (2020) 075}
  [\href{https://arxiv.org/abs/1910.06332}{{\ttfamily 1910.06332}}].

\bibitem{Agmon:2020pde}
N.~B. Agmon and Y.~Wang, \emph{{Classifying Superconformal Defects in Diverse
  Dimensions Part I: Superconformal Lines}},
  \href{https://arxiv.org/abs/2009.06650}{{\ttfamily 2009.06650}}.

\bibitem{Erickson:2000af}
J.~K. Erickson, G.~W. Semenoff and K.~Zarembo, \emph{{Wilson loops in N=4
  supersymmetric Yang-Mills theory}},
  \href{https://doi.org/10.1016/S0550-3213(00)00300-X}{\emph{Nucl. Phys. B}
  {\bfseries 582} (2000) 155}
  [\href{https://arxiv.org/abs/hep-th/0003055}{{\ttfamily hep-th/0003055}}].

\bibitem{Drukker:2000rr}
N.~Drukker and D.~J. Gross, \emph{{An Exact prediction of N=4 SUSYM theory for
  string theory}}, \href{https://doi.org/10.1063/1.1372177}{\emph{J. Math.
  Phys.} {\bfseries 42} (2001) 2896}
  [\href{https://arxiv.org/abs/hep-th/0010274}{{\ttfamily hep-th/0010274}}].

\bibitem{Pestun:2007rz}
V.~Pestun, \emph{{Localization of gauge theory on a four-sphere and
  supersymmetric Wilson loops}},
  \href{https://doi.org/10.1007/s00220-012-1485-0}{\emph{Commun. Math. Phys.}
  {\bfseries 313} (2012) 71} [\href{https://arxiv.org/abs/0712.2824}{{\ttfamily
  0712.2824}}].

\bibitem{Gomis:2008qa}
J.~Gomis, S.~Matsuura, T.~Okuda and D.~Trancanelli, \emph{{Wilson loop
  correlators at strong coupling: From matrices to bubbling geometries}},
  \href{https://doi.org/10.1088/1126-6708/2008/08/068}{\emph{JHEP} {\bfseries
  08} (2008) 068} [\href{https://arxiv.org/abs/0807.3330}{{\ttfamily
  0807.3330}}].

\bibitem{Ishibashi:1988kg}
N.~Ishibashi, \emph{{The Boundary and Crosscap States in Conformal Field
  Theories}}, \href{https://doi.org/10.1142/S0217732389000320}{\emph{Mod. Phys.
  Lett. A} {\bfseries 4} (1989) 251}.

\bibitem{Cardy:1989ir}
J.~L. Cardy, \emph{{Boundary Conditions, Fusion Rules and the Verlinde
  Formula}}, \href{https://doi.org/10.1016/0550-3213(89)90521-X}{\emph{Nucl.
  Phys. B} {\bfseries 324} (1989) 581}.

\bibitem{Komargodski:2011xv}
Z.~Komargodski, \emph{{The Constraints of Conformal Symmetry on RG Flows}},
  \href{https://doi.org/10.1007/JHEP07(2012)069}{\emph{JHEP} {\bfseries 07}
  (2012) 069} [\href{https://arxiv.org/abs/1112.4538}{{\ttfamily 1112.4538}}].

\bibitem{Luty:2012ww}
M.~A. Luty, J.~Polchinski and R.~Rattazzi, \emph{{The $a$-theorem and the
  Asymptotics of 4D Quantum Field Theory}},
  \href{https://doi.org/10.1007/JHEP01(2013)152}{\emph{JHEP} {\bfseries 01}
  (2013) 152} [\href{https://arxiv.org/abs/1204.5221}{{\ttfamily 1204.5221}}].

\bibitem{Lewkowycz:2013laa}
A.~Lewkowycz and J.~Maldacena, \emph{{Exact results for the entanglement
  entropy and the energy radiated by a quark}},
  \href{https://doi.org/10.1007/JHEP05(2014)025}{\emph{JHEP} {\bfseries 05}
  (2014) 025} [\href{https://arxiv.org/abs/1312.5682}{{\ttfamily 1312.5682}}].

\bibitem{Grover:2011fa}
T.~Grover, A.~M. Turner and A.~Vishwanath, \emph{{Entanglement Entropy of
  Gapped Phases and Topological Order in Three dimensions}},
  \href{https://doi.org/10.1103/PhysRevB.84.195120}{\emph{Phys. Rev. B}
  {\bfseries 84} (2011) 195120}
  [\href{https://arxiv.org/abs/1108.4038}{{\ttfamily 1108.4038}}].

\bibitem{Liu:2012eea}
H.~Liu and M.~Mezei, \emph{{A Refinement of entanglement entropy and the number
  of degrees of freedom}},
  \href{https://doi.org/10.1007/JHEP04(2013)162}{\emph{JHEP} {\bfseries 04}
  (2013) 162} [\href{https://arxiv.org/abs/1202.2070}{{\ttfamily 1202.2070}}].

\bibitem{Witten:1992qy}
E.~Witten, \emph{{On background independent open string field theory}},
  \href{https://doi.org/10.1103/PhysRevD.46.5467}{\emph{Phys. Rev. D}
  {\bfseries 46} (1992) 5467}
  [\href{https://arxiv.org/abs/hep-th/9208027}{{\ttfamily hep-th/9208027}}].

\bibitem{Witten:1992cr}
E.~Witten, \emph{{Some computations in background independent off-shell string
  theory}}, \href{https://doi.org/10.1103/PhysRevD.47.3405}{\emph{Phys. Rev. D}
  {\bfseries 47} (1993) 3405}
  [\href{https://arxiv.org/abs/hep-th/9210065}{{\ttfamily hep-th/9210065}}].

\bibitem{Shatashvili:1993kk}
S.~L. Shatashvili, \emph{{Comment on the background independent open string
  theory}}, \href{https://doi.org/10.1016/0370-2693(93)90537-R}{\emph{Phys.
  Lett. B} {\bfseries 311} (1993) 83}
  [\href{https://arxiv.org/abs/hep-th/9303143}{{\ttfamily hep-th/9303143}}].

\bibitem{Shatashvili:1993ps}
S.~L. Shatashvili, \emph{{On the problems with background independence in
  string theory}}, \href{https://doi.org/10.1007/3-540-58453-6_12}{\emph{Alg.
  Anal.} {\bfseries 6} (1994) 215}
  [\href{https://arxiv.org/abs/hep-th/9311177}{{\ttfamily hep-th/9311177}}].

\bibitem{Kutasov:2000qp}
D.~Kutasov, M.~Marino and G.~W. Moore, \emph{{Some exact results on tachyon
  condensation in string field theory}},
  \href{https://doi.org/10.1088/1126-6708/2000/10/045}{\emph{JHEP} {\bfseries
  10} (2000) 045} [\href{https://arxiv.org/abs/hep-th/0009148}{{\ttfamily
  hep-th/0009148}}].

\bibitem{Erdmenger:2013dpa}
J.~Erdmenger, C.~Hoyos, A.~O'Bannon and J.~Wu, \emph{{A Holographic Model of
  the Kondo Effect}},
  \href{https://doi.org/10.1007/JHEP12(2013)086}{\emph{JHEP} {\bfseries 12}
  (2013) 086} [\href{https://arxiv.org/abs/1310.3271}{{\ttfamily 1310.3271}}].

\bibitem{Oshikawa:1996dj}
M.~Oshikawa and I.~Affleck, \emph{{Boundary conformal field theory approach to
  the critical two-dimensional Ising model with a defect line}},
  \href{https://doi.org/10.1016/S0550-3213(97)00219-8}{\emph{Nucl. Phys. B}
  {\bfseries 495} (1997) 533}
  [\href{https://arxiv.org/abs/cond-mat/9612187}{{\ttfamily
  cond-mat/9612187}}].

\bibitem{Oshikawa:1996ww}
M.~Oshikawa and I.~Affleck, \emph{{Defect lines in the Ising model and boundary
  states on orbifolds}},
  \href{https://doi.org/10.1103/PhysRevLett.77.2604}{\emph{Phys. Rev. Lett.}
  {\bfseries 77} (1996) 2604}
  [\href{https://arxiv.org/abs/hep-th/9606177}{{\ttfamily hep-th/9606177}}].

\bibitem{Polchinski:2011im}
J.~Polchinski and J.~Sully, \emph{{Wilson Loop Renormalization Group Flows}},
  \href{https://doi.org/10.1007/JHEP10(2011)059}{\emph{JHEP} {\bfseries 10}
  (2011) 059} [\href{https://arxiv.org/abs/1104.5077}{{\ttfamily 1104.5077}}].

\bibitem{Konechny:2003yy}
A.~Konechny, \emph{{g function in perturbation theory}},
  \href{https://doi.org/10.1142/S0217751X04019469}{\emph{Int. J. Mod. Phys. A}
  {\bfseries 19} (2004) 2545}
  [\href{https://arxiv.org/abs/hep-th/0310258}{{\ttfamily hep-th/0310258}}].

\bibitem{Beccaria:2017rbe}
M.~Beccaria, S.~Giombi and A.~Tseytlin, \emph{{Non-supersymmetric Wilson loop
  in $ \mathcal{N} $ = 4 SYM and defect 1d CFT}},
  \href{https://doi.org/10.1007/JHEP03(2018)131}{\emph{JHEP} {\bfseries 03}
  (2018) 131} [\href{https://arxiv.org/abs/1712.06874}{{\ttfamily
  1712.06874}}].

\bibitem{Elitzur:1998va}
S.~Elitzur, E.~Rabinovici and G.~Sarkissian, \emph{{On least action D-branes}},
  \href{https://doi.org/10.1016/S0550-3213(98)00799-8}{\emph{Nucl. Phys. B}
  {\bfseries 541} (1999) 246}
  [\href{https://arxiv.org/abs/hep-th/9807161}{{\ttfamily hep-th/9807161}}].

\bibitem{Fredenhagen:2006dn}
S.~Fredenhagen, M.~R. Gaberdiel and C.~A. Keller, \emph{{Bulk induced boundary
  perturbations}}, \href{https://doi.org/10.1088/1751-8113/40/1/F03}{\emph{J.
  Phys. A} {\bfseries 40} (2007) F17}
  [\href{https://arxiv.org/abs/hep-th/0609034}{{\ttfamily hep-th/0609034}}].

\bibitem{Elitzur:2012wm}
S.~Elitzur, B.~Karni and E.~Rabinovici, \emph{{Induced Boundary Flow on the c =
  1 Orbifold Moduli Space}},
  \href{https://doi.org/10.1088/1751-8113/45/45/455401}{\emph{J. Phys. A}
  {\bfseries 45} (2012) 455401}
  [\href{https://arxiv.org/abs/1209.4344}{{\ttfamily 1209.4344}}].

\bibitem{Bianchi:2019umv}
L.~Bianchi, \emph{{Marginal deformations and defect anomalies}},
  \href{https://doi.org/10.1103/PhysRevD.100.126018}{\emph{Phys. Rev. D}
  {\bfseries 100} (2019) 126018}
  [\href{https://arxiv.org/abs/1907.06193}{{\ttfamily 1907.06193}}].

\bibitem{Herzog:2019rke}
C.~P. Herzog and I.~Shamir, \emph{{How a-type anomalies can depend on marginal
  couplings}},
  \href{https://doi.org/10.1103/PhysRevLett.124.011601}{\emph{Phys. Rev. Lett.}
  {\bfseries 124} (2020) 011601}
  [\href{https://arxiv.org/abs/1907.04952}{{\ttfamily 1907.04952}}].

\bibitem{Green:2007wr}
D.~R. Green, M.~Mulligan and D.~Starr, \emph{{Boundary Entropy Can Increase
  Under Bulk RG Flow}},
  \href{https://doi.org/10.1016/j.nuclphysb.2008.01.010}{\emph{Nucl. Phys. B}
  {\bfseries 798} (2008) 491}
  [\href{https://arxiv.org/abs/0710.4348}{{\ttfamily 0710.4348}}].

\end{thebibliography}\endgroup
\bibliographystyle{JHEP.bst}

\end{document}